\documentclass[11pt]{amsart}
\usepackage{latexsym,amssymb,amsmath,amscd,amsthm}
\numberwithin{equation}{section}
\theoremstyle{remark}
\newtheorem{theorem}{{\bf THEOREM}}[section]

\newtheorem{example}{{\bf EXAMPLE}}[section]
\newcommand{\bq}{\begin{equation}}
\newcommand{\bea}{\begin{array}}
\newcommand{\eea}{\end{array}}

\newcommand{\ga}{\alpha}
\newcommand{\gep}{\epsilon}
\newcommand{\gD}{\Delta}
\newcommand{\gl}{\lambda}

\newcommand{\gb}{\beta}
\newcommand{\ot}{\otimes}
\newcommand{\mf}{\mathfrak}
\newcommand{\mc}{\mathcal}

\newcommand{\dg}{\dagger}

\newcommand{\ul}[1]{\underline{#1}}

\newcommand{\go}{\omega}

\newcommand{\gG}{\Gamma}

\newcommand{\gs}{\sigma}

\newcommand{\gag}{\gamma}
\newcommand{\gd}{\delta}
\newcommand{\pp}{\partial}

\newcommand{\tl}{\tilde}
\newcommand{\na}{\nabla}
\newcommand{\gk}{\kappa}

\newcommand{\bl}{\blacklozenge}
\newcommand{\bs}{\blacksquare}

\newcommand{\bgs}{\bigstar}

\newcommand{\gS}{\Sigma}

\newcommand{{\DDD}}{D\!\!\!\!\!\!-}
\newcommand{\bx}{\Box}

\title{REMARKS ON PHOTONS AND THE AETHER}

\author{Robert Carroll\\University of Illinois, Urbana, IL 61801}

\date{July, 2005\thanks{email: rcarroll@math.uiuc.edu}}

\begin{document}

\bibliographystyle{plain}

\begin{abstract} 
We expand upon some topics reviewed and sketched in \cite{c1} with more details, embellishments,
and some new material of a speculative nature.
\end{abstract}

\maketitle

%\tableofcontents

\section{INTRODUCTION}
\renewcommand{\theequation}{1.\arabic{equation}}
\setcounter{equation}{0}

In the book \cite{c1} we discussed many aspects of e.g. electromagnetism (EM), the
aether, and the Schr\"odinger equation (SE) partly in connection with our study of the
quantum potential (QP).  We now want to examine further the nature of photons and
radiation in connection with a putative aether.  It is essential that we review some
of the background from \cite{c1} in order to motivate the aether treatment in Section 6.

\section{PHOTONS}
\renewcommand{\theequation}{2.\arabic{equation}}
\setcounter{equation}{0}

We begin with \cite{t1} (which is
also sketched in \cite{c1} in a somewhat different manner)
and in Section 5 give a related description following \cite{c5}.  We will also
examine further various points of view concerning the massless Klein-Gordon (KG)
equation, the SE, the Maxwell equations (ME), and the quantum vacuum.  For background
we mention here
\cite{a3,b2,c1,c3,d1,f6,f5,f3,f4,g1,g2,h4,h3,h5,h6,h7,h8,i1,i3,j1,k3,k2,k1,m1,n1,p1,p2,
p3,r1,r2,s1,t3}.
One takes massless photons as objects with energy E, momentum {\bf P}, and internal
angular momentum (or spin) {\bf S} with $E=c|{\bf P}|$ and ${\bf S}\times {\bf P}=0$.
It is presumed to have velocity c in the direction {\bf k} and to spin in a plane
perpendicular to {\bf k}, which is spanned by two vectors {\bf e} and {\bf b} where
\bq\label{2.1}
{\bf k}\cdot{\bf e}={\bf k}\cdot{\bf b}=0;\,\,{\bf k}\times{\bf e}={\bf b};\,\,{\bf
k}\times{\bf b}=-{\bf e};\,\,{\bf e}|=|{\bf b}|;\,\,{\bf e}\cdot{\bf b}=0
\end{equation}
One sets $\go=e=b$ (frequency) and $E=\hbar\go$ historically (with $|{\bf
S}|=\pm\hbar$) while $\gl=2\pi c/\go$ (which will eventually be identified with a wave
length).  The photon is considered as following a right of left handed helix generated
by the tip of {\bf e} where the plane of ${\bf e},\,{\bf b}$ moves along the direction
{\bf k} with velocity c.  These objects are exhibited via a photon tensor
\bq\label{2.2}
f^{\mu\nu}=\left(\begin{array}{cccc}
0 & e_1 & e_2 & e_3\\
-e_1 & 0 & b_3 & -b_2\\
-e_2 & -b_3 & 0 & b_1\\
-e_3 & b_2 & -b_1 & 0
\end{array}\right)
\end{equation}
which is $\ul{not}$ a field like the EM tensor $F^{\mu\nu}$ (see Section 5 for more
comments on the tensor nature of $f^{\mu\nu}$).
The dual tensor is
\bq\label{2.3}
f^{*\mu\nu}=\frac{1}{2}\gep^{\mu\nu\gs\rho}f_{\gs\rho}=
\left(\begin{array}{cccc}
0 & -b_1 & -b_2 & -b_3\\
b_1 & 0 & e_3 & -e_2\\
b_2 & -e_3 & 0 & e_1\\
b_3 & e_2 & -e_1 & 0
\end{array}\right)
\end{equation}
with $f^{\mu\nu}f_{\mu\nu}=2(e^2-b^2)=0$ and $f^{\mu\nu}f^*_{\mu\nu}=-4{\bf
e}\cdot{\bf b}=0$.  One works here in a Hilbert space $H=H^S\ot H^K$ with ${\bf S}\sim
{\bf S}\ot 1,\,\,{\bf P}\sim 1\ot {\bf P}$, and ${\bf R}\sim 1\ot {\bf R}$.  Now spin
is colinear with momentum (recall ${\bf S}\times{\bf P}=0$) and the spin eigenstates
$\chi_{\pm}$ correspond to helicities $\pm 1$ satisfying
\bq\label{2.4}
({\bf k}\cdot{\bf S})\chi_{\pm}=\pm\hbar\chi_{\pm}
\end{equation}
where {\bf k} is a unit vector in the direction of {\bf P}.  The spin operators will
be expressed via
\bq\label{2.5}
S_x=\hbar\left(\begin{array}{ccc}
0 & 0 & 0\\
0 & 0 & -i\\
0 & i & 0
\end{array}\right);\,\,S_y=\hbar\left(\begin{array}{ccc}
0 & 0 & i\\
0 & 0 & 0\\
-i & 0 & 0
\end{array}\right);\,\,S_z=\hbar\left(\begin{array}{ccc}
0 & -i & 0\\
i & 0 & 0\\
0 & 0 & 0
\end{array}\right)
\end{equation}
with $(S_j)_{k\ell}=-i\hbar\gep_{jk\ell}$.  One must distinguish here ${\bf k}\in H^K$
and ${\bf S}\in H^S$; the 2-dimensional spin space is orthogonal to {\bf k} with 
${\bf k}\cdot{\bf S}\sim\pm\hbar$ as indicated in \eqref{2.4}.  Now write $\psi_j\in
H^S\ot H^K$ with $j=1,2,3$ denoting components in $H^S$ and set ${\bf k}={\bf p}/|{\bf
p}|$.  An operator leaving invariant a photon state $\chi^k_{\pm}\ot\phi_p$ is
${\bf S}\cdot{\bf P}={\bf k}\cdot{\bf S}\ot|{\bf P}|$ where $|{\bf P}|\phi_p=
|{\bf p}|\phi_p$ (with $E=c|{\bf p}|=c|p|$) and one has
\bq\label{2.6}
{\bf S}\cdot{\bf P}\chi_{\pm}^k\ot\phi_p=\pm\frac{\hbar E}{c}\chi^k_{\pm}\ot \phi_p
\end{equation}
(the Hamiltonian is $H=(c/\hbar){\bf S}\cdot{\bf P}$ and a minus sign should be
interpreted as positive energy but negative helicity).  Then the time evolution of a
general photon state is
\bq\label{2.7}
i\hbar\pp_t\psi_j=(H)_{jk}\psi_k=\frac{c}{\hbar}({\bf S}\cdot{\bf P})_{jk}\psi_k
\end{equation}
(H is a $3\times 3$ matrix in $H^S$ whose components are operators in $H^K$).  Putting
${\bf P}=-i\hbar\na$ one has a SE for the photon, namely
\bq\label{2.8}
\frac{i}{c}\pp_t\psi_j(t,{\bf r})=-\gep_{jk\ell}\pp_{\ell}\psi_k(t,{\bf r})
\end{equation}
(since
$(c/\hbar)(-i\hbar\gep_{jk\ell})(-i\hbar\pp_{\ell})=-c\hbar\gep_{jk\ell}\pp_{\ell}$).
Note
$\hbar$ has disappeared and although this is a QM equation it does not have a classical
limit.
\\[3mm]\indent
{\bf REMARK 2.1.}
It is pointed out in \cite{t1} that there are conceptual errors in writing
$\psi_j=E_j+iB_j$ and deriving the Maxwell equations via $(i/c)\pp_t(E_j+iB_j)=
-\gep_{jk\ell}\pp_{\ell}(E_k+iB_k)$ in the form
\bq\label{2.9}
\frac{1}{c}\pp_tE_j=-\gep_{jk\ell}\pp_{\ell}B_k;\,\,\frac{1}{c}\pp_tB_j=\gep_{jk\ell}
\pp_{\ell}E_k
\end{equation}
(e.g. $(1/c)\pp_tE_1=-\gep_{123}\pp_3B_2-\gep_{132}\pp_2B_3=\pp_2B_3-\pp_3B_2$, 
etc. - note $\gep_{jk\ell}=-\gep_{j\ell k}$ in \cite{t1}).  The equations are correct
but the derivation is faulty since it identifies the 3-D space of states with the 
3-D physical space!$\hfill\bs$
\\[3mm]\indent
{\bf REMARK 2.2.}
Using the momentum representation one can write as in \cite{t1}
\bq\label{2.10}
\frac{\hbar}{c}\pp_t\psi_j(t,{\bf p})=\gep_{jk\ell}p_{\ell}\psi_k(t,{\bf p})
\end{equation}
but this is not $\vec{\psi}\times\vec{p}$ because the two vectors belong to different
spaces.  One can look also at stationary state solutions $\psi_j=exp[-(i/\hbar)Et)
\Phi_{j,E}$ where $({\bf k}={\bf p}/|{\bf p}|$)
\bq\label{2.11}
({\bf S}\cdot{\bf P})_{jk}\Phi_{k,E}=\frac{E\hbar}{c}\Phi_{j,E};\,\,\Phi_{j,E}({\bf
p})=
\chi_{\pm}^k\ot\gd\left(|{\bf p}|-\frac{E}{c}\right)
\end{equation}
For the corresponding position representation one would use $\chi^{k_0}_{\pm}\ot
\phi_{p_0}$ where
\bq\label{2.12}
\phi_{p_0}({\bf r})=\frac{1}{\sqrt{2\pi\hbar}^3}exp\left(\frac{i}{\hbar}{\bf
p}_0\cdot{\bf r}\right)
\end{equation}
(where ${\bf k}_0={\bf p}_0/|{\bf p}_0|$.$\hfill\bs$

\section{THE EM FIELDS}
\renewcommand{\theequation}{3.\arabic{equation}}
\setcounter{equation}{0}

In the last paper of \cite{t1} it is shown how to construct the EM fields from
knowledge of photons.  First one defines
\bq\label{3.1}
\gep_{+}=\frac{1}{\sqrt{2}}(\hat{{\bf e}}+i\hat{{\bf b}});\,\,\gep_{-}=
\frac{1}{\sqrt{2}}(i\hat{{\bf e}}+\hat{{\bf b}})
\end{equation}
where ${\bf e}=\go\hat{{\bf e}}$ and ${\bf b}=\go\hat{{\bf b}}$.  Then write
\bq\label{3.2}
{\bf e}_{+}(t)=\left(\frac{\go}{\sqrt{2}}\gep_{+}e^{-i\go t}+c.c.\right);\,\,{\bf
e}_{-}(t)=\left(\frac{\go}{\sqrt{2}}\gep_{-}e^{-i\go t}+c.c.\right)
\end{equation}
One writes ${\bf e}_s(t)=[(\go/\sqrt{2})\gep_sexp(-i\go t)+c.c.)$ and ${\bf b}_s(t)=
{\bf k}\times{\bf e}_s(t)$, uses momentum eigenfunctions as in Remark 2.2, and
writes $\phi_{s,p}=\chi^k_s\ot\phi_p$.  For a state with n photons having helicity
$s_j$ and momentum $p_j$ annihilation and creation operators are defined via
\bq\label{3.3}
a_s^{\dg}(p)\phi_{s_1p_1,\cdots,s_np_n}=\sqrt{n+1}\phi_{sp,s_1p_1,\cdots,s_np_n};
\end{equation}
$$a_s(p)\phi_{s_1p_1,\cdots,s_np_n}=\frac{1}{\sqrt{n}}\sum_1^n\gd_{s,s_i}\gd({\bf p}-
{\bf p}_i)\phi_{s_1p_1,\cdots,\widehat{s_ip_i},\cdots,s_np_n}$$
A vacuum state $\phi_0$ with zero photons is defined via $a_s(p)\phi_0=0$ and n-photon
states are built up via
\bq\label{3.4}
\phi_{s_1p_1,\cdots,s_np_n}=\frac{1}{\sqrt{n!}}a_{s_1}^{\dg}(p_1)\cdots
a_{s_n}^{\dg}(p_n)\phi_0
\end{equation}
A number operator is defined via $N_s(p)=a^{\dg}_s(p)a_s(p)$ and $N=\sum_s\int d^3p
N_s(p)$.  The total energy, momentum, and spin of a system of photons (each with
energy $E=c|p|=\hbar\go$ and spin $\pm\hbar$) is then
\bq\label{3.5}
H=\sum_s\int d^3p\hbar\go N_s(p);\,\,{\bf P}=\sum_s\int d^3p{\bf p}N_s(p);
\end{equation}
$${\bf S}=\int d^3p\hbar{\bf k}(N_{+}(p)-N_{-}(p))$$
One defines then Hermitian operators
\bq\label{3.6}
{\bf E}({\bf r},t)=\frac{1}{2\pi\hbar}\sum_s\int
d^3p\sqrt{\go}\left(ia_s(p)\gep_se^{(i/\hbar)({\bf p}\cdot{\bf r}-Et)}+h.c.\right);
\end{equation}
$${\bf B}({\bf r},t)=\frac{1}{2\pi\hbar}\sum_s\int d^3p\sqrt{\go}\left(ia_s(p)({\bf
k}\times{\bf \gep}_s)e^{(i/\hbar)({\bf p}\cdot{\bf r}-Et)}+h.c.\right);$$
$${\bf A}({\bf r},t)=\frac{c}{2\pi\hbar}\sum_s\int d^3p\frac{1}{\sqrt{\go}}
\left(a_s(p)\gep_se^{(i/\hbar)({\bf p}\cdot{\bf r}-Et)}+h.c.\right)$$
Then 
\bq\label{3.7}
{\bf E}=-\frac{1}{c}\pp_t{\bf A};\,\,{\bf B}=\na\times{\bf A};\,\,
H=\frac{1}{8\pi}\int d^3r({\bf E}^2+{\bf B}^2);
\end{equation}
$${\bf P}=\frac{1}{8\pi c}\int d^3r({\bf E}\times {\bf B}-{\bf B}\times {\bf E};\,\,
{\bf S}=\frac{1}{8\pi c}\int d^3r({\bf E}\times{\bf A}-{\bf A}\times{\bf E})$$
and one checks the Maxwell equations
\bq\label{3.8}
-\na\times{\bf E}=\frac{1}{c}\pp_t{\bf B};\,\,\na\times{\bf B}=\frac{1}{c}\pp_t{\bf
E};\,\,\na\cdot{\bf E}=\na\cdot{\bf B}=0
\end{equation}
Thus photons are posited as the fundamental objects and they generate EM fields as a
collective manifestation.
\\[3mm]\indent
Next one defines the ``singular" function (cf. \cite{t1} for details)
\bq\label{3.9}
D(\vec{\rho},\tau)=\frac{-1}{(2\pi\hbar)^3}\int d^3pe^{(i/\hbar){\bf p}\cdot\vec{\rho}}
\frac{Sin(\go\tau)}{\go}=
\end{equation}
$$=\frac{-1}{8\pi^2c\rho}[\gd(\rho-c\tau)-\gd(\rho+c\tau)]$$
Here $\rho=|\vec{\rho}|$ where $\vec{\rho}\sim {\bf r}_1-{\bf r}_2$ and one can
say that $D(\vec{\rho},\tau)$ has support on the light cone (cf. also \cite{m1}).  
This leads to
\bq\label{3.10}
[E_i({\bf r}_1,t_1),E_j({\bf r}_2,t_2)]=-4\pi i\hbar c^2\left(\frac{\gd_{ij}}{c^2}
\pp_{t_1}\pp_{t_2}+\pp_{r_1,i}\pp_{r_2,j}\right)D({\bf r}_1-{\bf r}_2,t_1-t_2)
\end{equation}
$$[B_i({\bf r}_1,t_1),B_j({|bf r}_2,t_2)]=-4\pi i\hbar  c^2\left(\frac{\gd_{ij}}{c^2}
\pp_{t_1}\pp_{t_2}+\pp_{r_1,i}\pp_{r_2,j}\right)D({\bf r}_1-{\bf r}_2,t_1-t_2)$$
$$[E_i({\bf r}_1,t_1),B_j({\bf r}_2,t_2)]=4\pi i\hbar c\gep_{ijk}\pp_{t_1}\pp_{r_1,k}
D({\bf r}_1-{\bf r}_2,t_2-t_1)$$
Note that the singular nature of D is really unacceptable in QM (e.g. because of the
uncertainty principle) and one could conclude that the field strengths are not
measurable quantities (cf. the first paper in \cite{t1}).  On the other hand field
averages can be accepted in QM.  This is one feature leading to the approch in
\cite{t1} based on the photon as fundamental.  The EM fields are considered essentially
as a classical macroscopic ideas and are not ``basic".  Such an argument might be
extendable quite generally to cast suspicion on many results involving singular
behavior or generalized solutions of partial differential equations (distributions).
The ``classical" theory might require e.g. averaging of dependent variables or some
new physics (not necessarily QM) in order to retain any meaning.
\\[3mm]\indent
One looks next at the expectation values of fields in the quantum state describing 
a system of photons.  For the vacuum described via $\phi_0$ one has
\bq\label{3.11}
<\phi_0,E({\bf r},t)\phi_0>=<\phi_0,{\bf B}({\bf r},t)\phi_0>=0
\end{equation}
as expected.  However one can show that e.g.
\bq\label{3.12}
<\phi_0,{\bf E}^2({\bf r},t)\phi_0>=\frac{2}{(2\pi\hbar)^2}
\int d^3p\,\go
\end{equation}
indicating that there are fluctuations of the electric field in vacuum.
For a quantum state of n photons in the same state with fixed helicity and momentum
one has (cf. \cite{t1})
\bq\label{3.13}
\phi=\phi_{n(s_1p_1)}=\frac{1}{\sqrt{n!}}(a^{\dg}_{s_1}({\bf p}_1))^n\phi_0;\,\,
<\phi,{\bf E}({\bf r},t)\phi>=<\phi,{\bf B}({\bf r},t)\phi>=0
\end{equation}
which is somewhat strange.  However for an indefinite number of photons in a
superposition of states $\psi=\sum_nC_n\phi_{n(s_1p_1)}$ one has
\bq\label{3.14}
<\psi,{\bf E}({\bf
r},t)\psi>=\frac{\sqrt{\go_1}}{2\pi\hbar}\left(i\sum_nC_n^*C_{n+1}
\gep_{s_1}e^{(i/\hbar)({\bf p}_1\cdot{\bf r}-E_1t)}+c.c.\right)
\end{equation}
One concludes here that the EM field of an indefinite number of photons all with the
same helicity and momentum is a plane wave with circular polarization.  The quantum
state where all photons are in the same one photon state of fixed helicity and
momentum is a Bose-Einstein condensate (?).
\\[3mm]
{\bf REMARK 2.1.}
We extract here from \cite{n1} for a few philosophical observations.  The photon, as
an elementary ``particle" is unique; it is the only elementary particle of energy.
A relativistic energy equation should be $E^2=p^2c^2+m_0c^2=p^2c^2$
since the rest mass $m_0=0$.  In the frame of the moving photon the photon's 
energy is stored as rotational (spin) energy where $E=\hbar\nu=\hbar c/\gl$ with
$\nu$ the frequency and $\gl$ the wave length.  Hence the greater the energy the
smaller the wave length and one expects to find a lower bound for the wavelength.
For a ``particle" the angular momentum is $L=mrw$ limited by $L=mrc$ and replacing L
by the spin S one has $(\spadesuit)\,\,\hbar=mrc$ where m is a putative mass presumably
``generated" by the spin (see here also \cite{t1} for toy models with extended energy
distributions).  Assume the concept of Schwartzschild radius R is valid for the photon
where for a black hole $R=2Gm/c^2$ or $(\clubsuit)\,\,(R
/m)=(2G/c^2)$.  The right side
of $(\clubsuit)$ is a constant but for the photon the radius decreases as the ``mass"
increases; hence there is a unique value of radius and mass for which a photon can
behave as a black hole.  Combining $(\clubsuit)$ with $(\spadesuit)$ one finds 
$(\bl)\,\,m=\sqrt{\hbar c/2G}$ for the Planck mass, which here is the maximum
``pseudomass" permitted for the photon.  This corresponds to a maximum energy
of $mc^2=(\sqrt{\hbar c/2G})c^2=8.61\times 10^{22}$ MeV and the highest energy so far
observed for a photon is apparently less than this.  It is suggested that pair
production or photon ``splitting" will ensue at the energy limit.$\hfill\bs$

\section{THE ZERO POINT FIELD - ZPF}
\renewcommand{\theequation}{4.\arabic{equation}}
\setcounter{equation}{0}

This is a murky subject and essentially involves understanding the quantum vacuum,
which of course still retains some mysteries.  We gave some hesitant and heuristic
comments on ZPF in \cite{c1}, based on
\cite{b2,d1,h5,h6,h7,h8,i1,i3,m1,p1,p2,p3,r1,r2}, which upon hindsight seem woefully
inadequate.  Some of this is also summarized and enhanced in a recent paper \cite{r1}
(first paper).  However we go here to the lovely collection
of papers by J. Field (see e.g. \cite{f7,f8}) for an aper\c{c}u of basic physical
connections between QM, thermodynamics, and special relativity.  This will serve as a
complement to Sections 2-3.  We begin with \cite{f7} (first paper) which in a sense
follows the spirit of Feynman's QED where the fundamental concepts of QM are explained
in terms of the interactions of photons and electrons.  One recalls first the energy
momentum vector $P=m(dX/d\tau)\sim ((E/c),p_x,p_y,p_z)$ with $X=(ct,x,y,z)$
and $\tau$ the proper time (time observed in the rest frame).  If the inertial frame
S' is moving with uniform velocity $\gb c$ relative to the frame S along the common
$x,\,x'$ axis with $0y$ parallel to $0y'$ then the 4-vectors as observed in S, S' are
related by Lorentz transform (LT) equations
\bq\label{4.1}
p'_x=\gag(p_x-\gb p_t);\,\,p'_y=p_y;\,\,p'_z=p_z;\,\,p'_t=\gag(p_t-\gb p_x);
\end{equation}
$$\gag=\frac{1}{\sqrt{1-\gb^2}};\,\,p_t=\frac{E}{c}$$
As $m\to 0$ P is still well defined so one has an energy momentum vector say $P_{\gag}
=[(E_{\gag}/c),(E_{\gag}/c)Cos(\phi),(E_{\gag}/c)Sin(\phi),0]$ for a photon of energy 
$E_{\gag}$  moving in the $(x,y)$ plane in a direction making an angle $\phi$ with the 
$x$ axis.  A plane EM wave will be associated with a large number of photons in general
and for such a collection, all with the same 4-vector $P_{\gag}$, one finds from the 
LT equations an EM wave with
\bq\label{4.2}
\nu'=\nu\gag(1-\gb Cos(\phi));\,\,E'_T=E_T\gag(1-\gb Cos(\phi))
\end{equation}
(here $\nu\sim$ frequency and $E_T\sim$ total energy).  Using \eqref{4.1} the energies
of the photons in the EM wave transform via $E'_{\gag}=E_{\gag}\gag(1-\gb Cos(\phi))$.
If $n_{\gag}$ is the total number of photons then $E_T=n_{\gag}E_{\gag}$ which yields
\eqref{4.2}.  Further one sees immediately that
$E_{\gag}/\nu=E'_{\gag}/\nu'=constant$ and calling this constant $\hbar$ one finds that
$E_{\gag}=\hbar\nu$ which identifies $\hbar$ with Planck's constant.  Consequently
Planck's constant arises from consistency between the relativistic kinematics 
of photons, considered to be massless particles, and the relativistic Doppler effect
for classical EM waves.  Note also that using $\gl=c/\nu$ and $E_{\gag}=\hbar\nu
=p_{\gag}c$ one arrives at the deBroglie relation $p_{\gag}=\hbar/\gl=\hbar\nu/c$.
\\[3mm]\indent
Now from the energy density of a plane EM wave, namely
\bq\label{4.3}
\rho_W=\frac{{\bf E}^2+{\bf B}^2}{8\pi}
\end{equation}
the photon interpretation gives immediately Poynting's formula for the energy flow
F per unit area per unit time, namely $F=c\rho_W$ as well as the formula for the
radiation pressure $P_{rad}$ of a plane wave at normal incidence on a perfect
reflector, namely $P_{rad}=2\rho_W$.  Note the number of photons incident is
$F/E_{\gag}$ per unit area per unit time so the total momentum transferred is then
$p_{\gag}(F/E_{\gag})=F/c=\rho_W$; but a perfect reflector will not absorb energy so
an equal number of photons are re-emitted, yielding the factor of 2.  
Now consider a plane EM wave of wavelength
$\gl$ moving in free space parallel to the positive $x$ direction in the frame S,
written as
\bq\label{4.4}
E_y=E_0e^{\Phi};\,\,H_z=H_0e^{\Phi};\,\,\Phi=2\pi i\frac{(x-ct)}{\gl};\,\,E_0=H_0=A
\end{equation}
The time averaged energy density per unit volume $\bar{\rho}_W$ is $\bar{\rho}_W=
(E_0^2/8\pi)=(H_0^2/8\pi)=(A^2/8\pi)$.  Assuming that the wave consists of a beam of
photons of energy $\hbar\nu$ the average number density of photons $\bar{\rho}_{\gag}$
in the wave is $\bar{\rho}_W/\hbar\nu$ so one gets
$\bar{\rho}_{\gag}=A^2/8\pi\hbar\nu$.  This is the point where one now leaps across the
chasm separating the classical and quantum worlds.  First the use of a complex
exponential to represent a classical EM wave is convenient but it is really the real or
imaginary parts that come into play (Cosines and Sines); for QM the complex exponential
is mandatory.  Second one uses the definition of wavelength together with
$E_{\gag}=\hbar\nu$ and $p_{\gag}=\hbar/\gl$ to replace in the complex exponential
the wave parameter $\gl$ by the particle parameters $E_{\gag}$ and $p_{\gag}$.
The parameter c is part of both descriptions (photons and EM waves) and this leads to 
the complex exponential describing photons in the form
\bq\label{4.5}
u_p=u_0exp\left[\frac{2\pi i}{\hbar}(p_{\gag}x-E_{\gag}t)\right]=exp\left[\frac
{2\pi i}{\hbar}(P\cdot X)\right];\,\,u_0=\frac{A}{\sqrt{8\pi E_{\gag}}}
\end{equation}
In this situation $\bar{\rho}_{\gag}=|u_p|^2=u_0^2$ and for the case of very weak EM
fields such that $\bar{\rho}_{\gag}<< 1$ it follows that $|u_p|^2dV$ can be thought
of as the probability that a photon is in the volume $dV$; for large numbers of photons
or strong EM fields this probabilistic interpretation is not appropriate.
Note also that one can write
\bq\label{4.6}
{\mc P}_x=-i\frac{\hbar}{2\pi}\frac{\pp}{\pp x};\,\,{\mc E}=i\frac{\hbar}{2\pi}\frac
{\pp}{\pp t};\,\,{\mc P}_xu_p=p_{\gag}u_p;\,\,{\mc E}u_p=E_{\gag}u_p
\end{equation}
Note also for $f$ an arbitrary function on space-time
\bq\label{4.7}
{\mc P}_x(xf)=-\frac{i\hbar}{2\pi}\frac{\pp(xf)}{\pp x}=-\frac{i\hbar}{2\pi}f+x{\mc
P}_xf
\end{equation}
Repeated use of \eqref{4.6}, \eqref{4.7}, etc. and the relation $E_{\gag}=p_{\gag}c$
gives
\bq\label{4.8}
(c^2{\mc P}_x^2-{\mc E}^2)u_p=\left(\frac{\hbar c}{2\pi}\right)\bx u_p=0;\,\,
\bx=\frac{1}{c^2}\frac{\pp^2}{\pp t^2}-\frac{\pp^2}{\pp x^2}
\end{equation}
so $u_p$ will satisfy the Maxwell-Lorentz equation $\bx u_P=0$.  If one uses
instead of $E_{\gag}=p_{\gag}c$ the general energy momentum relation for massive
particles $E^2=p^2c^2+m^2c^4$ one can arrive at the Klein-Gordon (KG) equation and 
expanding $E\simeq mc^2+(p^2/2m)+\cdots$ the Schr\"odinger equation (SE) will 
result.
\\[3mm]\indent
Consider next ($p_{\gag}\to\hbar/\gl$ and $E_{\gag}\to \hbar c/\gl$)
\bq\label{4.9}
\chi=\frac{1}{2}(u_p+u_p^*)=\Re(u_p)=u_0Cos\left[\frac{2\pi}{\hbar}(p_{\gag}x-E_{\gag}
t)\right]\to u_0Cos\left(\frac{2\pi(x-ct)}{\gl}\right)
\end{equation}
This equation is then a bridge back across the chasm from QM to the classical world
(cf. \eqref{4.5}).  Just as the quantum wave function is only meaningful in the limit
of very low photon density so the function $\chi$ is meaningful only in the limit of
high photon density.  $\chi$ is not an eigenfunction of either $E_{\gag}$ or $p_{\gag}$
and is a real function.  The time average of $\chi^2$ is $1/2$ the mean photon density
$\bar{\rho}_{\gag}$ and $\bar{\rho}_W=\hbar\nu\bar{\rho}_{\gag}$.  In a typical
situation $\bar{\rho}_{\gag}\gD V$ is much larger than 1 and no probabilistic meaning
can be attached to it.
\\[3mm]\indent
We show next following \cite{f7} how to derive the Maxwell equations using only
Coulomb's inverse square law, special relativity, and Hamilton's principle.  Thus
take two objects $O_i$ of masses $m_i$ and electric charges $q_i$ with no external
forces.  The spatial distance separating them in the common center of mass frame
is ${\bf x}_{12}={\bf x}_1-{\bf x}_2$.  One constructs a most general Lorentz invariant
Lagrangian in a nonrelativistic reference frame via ($x_i\sim{\bf
x}_i$)
\bq\label{4.10}
L(x_1,u_1,x_2,u_2)=-\frac{m_1u_1^2}{2}-\frac{m_2u_2^2}{2}-\frac{j_1\cdot
j_2}{c^2\sqrt{-(x_1-x_2)^2}}
\end{equation}
where the $j_i=q_1u_i$ are current 4-vectors; this is then put into the machinery of
Hamilton's principle so that
\bq\label{4.11}
\frac{d}{d\tau}\left(\frac{\pp L}{\pp u_i^{\mu}}\right)-\frac{\pp L}{\pp
x_i^{\mu}}=0;\,\,(i=1,2;\,\,\mu=1,2,3,4)
\end{equation}
Since the Lagrangian is a Lorentz scalar this provides a description of the motion of 
the $O_i$ in any inertial reference frame.  Note that if one introduces a 4-vector
potential ${\bf A}_2={\bf j}_2/cr_{12},\,\,r_{12}=|{\bf r}_{12}|$ the standard Lorentz
invariant Lagrangian, describing the motion of $O_1$ in the EM field created by $O_2$,
namely
$L(x_1,u_1)=-(m_1u_1^2/2)-(1/c)q_1{\bf u}_1\cdot{\bf A}_2$, is recovered (and similarly
for motion of $O_2$ in the field of $O_1$).  Now write $\pp_i=-\pp^1\equiv(\pp/\pp
x^i)\equiv\na_i$ and set ${\bf p}=m{\bf u}$ along with
\bq\label{4.12} 
E^i=\pp^iA^0-\frac{1}{c}\frac{\pp A^i}{\pp t}=\pp^iA^0-\pp^0A^i;\,\,B^k=-\gep_{ijk}
(\pp^iA^j-\pp^jA^i)=(\na\times{\bf A})^k
\end{equation}
Some calculation (cf. \cite{f7})
gives then the 3-D Lorentz force equation and a relativistic Biot-Savart Law in the
form
\bq\label{4.13}
\frac{d{\bf p}}{dt}=q\left[{\bf E}+\frac{{\bf v}}{c}\times{\bf B}\right];\,\,
{\bf B}=\frac{q_2\gag_2({\bf v}_2\times{\bf r})}{cr^3}=\frac{{\bf j}\times{\bf r}}
{cr^3};
\end{equation}
$${\bf E}=\frac{j_2^0{\bf r}}{cr^3}-\frac{1}{c^2r}\frac{d{\bf j}_2}{dt}-
\frac{{\bf j}_2}{c^2}\frac{({\bf r}\cdot{\bf v}_2)}{r^3}$$
where ${\bf r}={\bf r}_{12}$.  
The Maxwell equations can be derived immediately from \eqref{4.12} along with the
Faraday-Lenz law, Ampere's law, etc. (cf. \cite{f7} for details).
\\[3mm]\indent
Now concerning the ZPF we collect some background information as follows.
\begin{enumerate}
\item
It seems well established that there is a unique Lorentz invariant spectral energy
density in the EM vacuum of the form $\rho(\go)=\rho_0(\go)=\hbar \go^2/2\pi^2c^3$
(cf. \cite{b2,m1}).  An observer moving with constant velocity in the EM vacuum
perceives no force.
\item
Following \cite{d2,u1} an object undergoing uniform constant acceleration $a$ in the
vacuum perceives himself to be immersed in a thermal bath at temperature $T=\hbar
a/2\pi kc$ ($k\sim$ Boltzman constant).
\item
One recalls also that there is a zero point energy $(1/2)\hbar\go$ attached to a
quantum harmonic oscillator.  Also since there are $(\go^2/2\pi^2c^3)d\go$ field nodes
per unit volume in the frequency interval $[\go,\go+d\go]$ one obtains the spectral
density $\rho_0(\go)=\hbar\go^4/2\pi^2c^3$ of Item 1 (cf. \cite{m1}).
\item
In \cite{b2} one derives the Planck radiation law for the blackbody spectrum without
the formalism of quantum theory.  It is assumed only that (i) There is classical,
homogeneous, and fluctuating EM EM radiation at absolute zero with Lorentz invariant
spectrum.  (ii) Classical EM theory holds for a dipole oscillator. (iii) A free
particle in equilibrium with blackbody radiation has classical kinetic energy $(1/2)kT$
per degree of freedom.  
This leads then to the zero point energy density shown above and to Planck's formula
\bq\label{4.14}
\rho(\go,T)=\frac{\go^2}{\pi^2c^3}\left[\frac{\hbar\go}{exp[(\hbar\go/kT)-1]}+\frac
{1}{2}\hbar\go\right]
\end{equation}
If the zero point energy is ignored one obtains the Rayleigh-Jeans formula
\bq\label{4.15}
\rho(\go,T)=\left(\frac{\go^2}{\pi^2c^3}\right)kT
\end{equation}
Here (the quantum number) $\hbar$ arises in \eqref{4.14} as a linear factor in
calculating the Lorentz invariant spectral density and can later be identified with
Planck's constant (so the derivation is classical).
\item
Going again to \cite{b2} one finds a lovely discussion involving entropy and energy
fluctuations following and modifying Einstein's arguments.  Thus one considers a 
cavity containing thermal radiation separated into large and small volumes V and
${\mc V}$.  The energy ${\mc U}$ of EM radiantion in ${\mc V}$ between frequencies
$\go$ and $\go+d\go$ undergoes spontaneous fluctuations creating a change in the
corresponding entropy.  Let $\gS$ (resp. ${\mf S}$) be the entropy contributed between
$\go$ and $\go+d\go$ for V (resp. (${\mc V}$).  Then for $\gep$ the entropy
fluctuation in  ${\mc V}$
\bq\label{4.16}
S(\gep)=\gS+{\mf S}=\gS_0+{\mf S}_0+(\pp_{\gep}\gS+\pp_{\gep}{\mf S})\gep+
\frac{1}{2}\left(\frac{\pp^2\gS}{\pp\gep^2}+\frac{\pp^2{\mf S}}{\pp\gep^2}\right)
\gep^2+\cdots
\end{equation}
where $\gS_0,\,\,{\mf S}_0$ signify equilibrium entropies where the fluctuation is
zero.  The first derivatives vanish at $\gep=0$ and if $V>>{\mc V}$ one finds
$S(\gep)\simeq\gS_0+{\mf S}_0+(1/2)(\pp^2{\mf S}/\pp{\mc U}^2)\gep^2$.
Now there is probabilistic entropy $(\clubsuit)\,\,S_{prob}=(S_{prob})_0+klog(W)$
(or $W=cexp(S_{prob}/k)$) where W is the number of microstates giving the same
macrostate.  There is also caloric entropy $S_{cal}$ where $dS_{cal}=dQ/T$ for
reversible processes.  Then write
\bq\label{4.17}
dW=cexp\left[\frac{S_{prob}}{k}\right]d\gep=\hat{c}exp\left[\frac{1}{2k}\frac
{\pp^2S_{prob}}{\pp{\mc U}^2}\gep^2\right]d\gep
\end{equation}
Some classical argument (cf. \cite{b2}, paper 2) involving $<\gep^2>=\int\gep^2dW\sim
(\pi^2c^3/\go^2)\rho^2d\go,\,\,{\mc U}=\rho d\go,$ and $\pp^2{\mf S}_{prob}/\pp{\mc
U}^2=-k/<\gep^2>$ leads then to $(\spadesuit)\,\,\pp^2S_{prob}/\pp E^2=-(k/E^2)$ 
for average oscillator energy
E.  Note in fact directly from the definition $(\clubsuit)$ one has $\pp S_{prob}/\pp
E=k/E$ leading to $(\spadesuit)$.  Now Einstein assumed that $S_{prob}=S_{cal}$ in
$(\spadesuit)$ and produced $E=kT$ along with the Planck formula (cf. \eqref{4.14})
$E=\hbar\go/[exp(\hbar\go/kT)-1]$ (with the zero point term missing).  Note here 
(using (4.14)) that the average energy of an oscillator is
\bq\label{4.18}
<\gep>=\frac{\pi^2c^3}{\go^2}\rho(\go,T)=\frac{\hbar\go}{exp(\hbar
\go/kT)-1}+\frac{1}{2}\hbar\go=E
\end{equation}
Now Boyer modifies Einstein's argument in a way which recovers the zero point term
(and (4.18)).
Indeed he writes $<\gep^2>=<\gep^2>_{ZPF}+<\gep^2>_{cal}$ and finds that 
\bq\label{4.19}
\frac{\pp^2S_{cal}}{\pp E^2}=-\frac{k}{E^2-(\hbar\go/2)^2}
\end{equation}
leading to \eqref{4.18}.
\end{enumerate}

\section{MORE ON PHOTONS}
\renewcommand{\theequation}{5.\arabic{equation}}
\setcounter{equation}{0}

We go here to \cite{b5,b6,b4,c5,g3,h9,h10,m6,s2} for some interesting developments
concerning the localization of photons and their structure.  One knows of course that
the methods of quantum field theory (QFT) work for a description of photon activity
but we want to examine more direct connections to EM fields, Maxwell's equations,
and wave-particle duality.  First from \cite{s2} one argues that a photon wave
function can be introduced if one is willing to redefine in a physically meaningful
manner what one wishes to mean by such a wave function.  First one introduces a naive
single photon wave function.  Then 
one produces a second
quantized many photon theory approached via many particle physics
(which will correspond to the quantization of the free radiation
field) and then recovers the naive single photon wave function by looking at
the manifold of one photon states.  There are apparently connections to the work of
\cite{b4} to which we don't have access at the moment.
\\[3mm]\indent
Now photons can be of positive or negative helicity and being massless one has
$E=cp$ where $p=|{\bf p}|$.  If one introduces probability amplitudes for photons of
momentum {\bf p} and helicity $\pm$, namely $\gag_{\pm}({\bf p},t)$ which would be
expected to satisfy a Schr\"odinger type equation $(\bl)\,\,i\hbar\pp_t\gag_{\pm}({\bf
p},t)=cp\gag_{\pm}({\bf p},t)$.  Next for each {\bf p} introduce two unit vectors
$\hat{{\bf e}}_i(\hat{{\bf p}})$ where $\hat{{\bf p}}={\bf p}/|{\bf p}|$ such that
$\hat{{\bf e}}_1,\hat{{\bf p}}_2,\hat{{\bf p}}$ form a right handed triad (cf. here
Section 1).  Then define helicity vectors ${\bf e}_{\pm}(\hat{{\bf
p}})=\mp(1/\sqrt{2})[\hat{{\bf e}}_1\pm i\hat{{\bf e}}_2]$ and write
$\vec{\gag}_{\pm}={\bf e}_{\pm}\gag_{\pm}$ with
\bq\label{5.1}
\vec{\gag}^*_{+}\cdot\vec{\gag}_{+}d{\bf p}=\gag^*_{+}\gag_{+}d{\bf p}
\end{equation}
for the probability of detecting a photon of positive helicity and momentum {\bf p}
between {\bf p} and ${\bf p}+d{\bf p}$ (similarly for negative helicity).  Note also
that $\vec{\gag}^*_{+}\cdot\vec{\gag}_{-}=0$.  Then define Fourier transforms
\bq\label{5.2}
\Phi_{\pm}({\bf r},t)=\int\frac{d{\bf p}}{(2\pi\hbar)^{3/2}}\gag_{\pm}({\bf
p},t)e^{i{\bf p}\cdot{\bf r}/\hbar}
\end{equation}
One then checks that $(\bl)$ is satisfied if
\bq\label{5.3}
i\hbar\pp_t\Phi_{\pm}({\bf r},t)=\pm c\hbar\na\times\Phi_{\pm}({\bf r},t)
\end{equation}
From the assumption that one is dealing with a single photon there results
\bq\label{5.4}
\int [\vec{\gag}^*_{+}\cdot\vec{\gag}_{+}+\vec{\gag}^*_{-}\cdot\vec{\gag}_{-}]d{\bf p}
=1\equiv\int[\Phi^*_{+}\cdot\Phi_{+}+\Phi^*_{-}\cdot\Phi_{-}]d{\bf r}=1
\end{equation}
The dynamical equations $(\bl)$ or \eqref{5.3} guarantee that if these equations
\eqref{5.4} are true at one time then they are satisfied at all later times.
In fact there results
\bq\label{5.5}
\Phi_{\pm}({\bf r},t)=\int\frac{d{\bf p}}{(2\pi\hbar)^{3/2}}\vec{\gag}_{\pm}({\bf
p},0)e^{-icpt/\hbar}e^{i{\bf p}\cdot{\bf r}/\hbar}
\end{equation}
There is then a temptation to try and identify the $\Phi_{\pm}$ as position
representation probability amplitudes for photons of positive or negative helicity
or perhaps their sum $\Phi_{+}+\Phi_{-}$ as a position representation of a photon.
However photons are not localizable so this doesn't work.  One way around (cf. \cite
{j2}) is to show that an operator representating the number of photons in an arbitrary
volume V can be defined but not as the integral over V of a photon density operator.
Another approach (cf. \cite{m7}) is to determine an operator representing the number
of photons in a volume V as the integral over V of a so called detection operator which
(when the linear dimensions of V are large compared to the photon wavelengths) leads
to a simple formula for the probability that n photons are present in V (cf. also 
\cite{c5} and Section 5.1 for coarse grained photon density and current density operators).  Here
one proceeds differently following \cite{s2} and looks for a probability amplitude
for the photon energy to be detected about $d{\bf r}$ of ${\bf r}$ in the form
$\Psi^*\cdot\Psi d{\bf r}$ with normalizations in the sense that
\bq\label{5.6}
\int\Psi^*\cdot\Psi d{\bf r}=\int cp[\vec{\gag}^*_{+}\cdot\vec{\gag}_{+}+
\vec{\gag}^*_{-}\cdot\vec{\gag}_{-}]d{\bf p}
\end{equation}
To do this one sets $\Psi=\Psi_{+}+\Psi_{-}$ with
\bq\label{5.7}
\Psi_{\pm}({\bf r},t)=\int\frac{\sqrt{cp}d{\bf
p}}{(2\pi\hbar)^{3/2}}\vec{\gag}_{\pm}({\bf p},t)e^{i{\bf p}\cdot{\bf r}/\hbar}
\end{equation}
One notes then that $i\hbar\pp_t\Psi_{\pm}=\pm c\hbar\na\times\Psi_{\pm}$ and
one must satisfy (cf. \eqref{5.5}) an initial condition given by the $t=0$ case of
\bq\label{5.8}
\Psi_{\pm}({\bf r},t)=\int\frac{\sqrt{cp}d{\bf
p}}{(2\pi\hbar)^{3/2}}\vec{\gag}_{\pm}({\bf p},0)e^{-icpt/\hbar}e^{i{\bf p}\cdot{\bf
r}/\hbar}
\end{equation}
Note here that {\bf p} (the usual photon momentum) and {\bf r}, the position associated
with the photon energy, are not conjugate variables.
One then builds up a QFT of of the free radiation field via many particle physics
(not from a canonical formulation of the EM fields) and this is equivalent to standard
canonical quantization.  Moreover upon specializing to one photon the energy 
functions $\Psi_{\pm}$ above are recovered.  Thus it is reasonable to describe the
single photon energy distribution in a region $d{\bf r}$ about {\bf r} via
$\Psi^*({\bf r},t)\cdot\Psi({\bf r},t)d{\bf r}$.  Further it is shown that in a
spontaneous emission process the wave function $\Psi({\bf r},t)$ generated is a causal
field, propagating out from the emitting atom at the speed of light.
\\[3mm]\indent
Now one goes to \cite{h9,h10} where the Bohr photon having a specific size and shape
is discussed.  This involves a circularly polarized photon being a monochromatic
EM traveling wave confined within a circular ellipsoid of length equal to the
wavelength ($\gl$) and diameter $\gl/\pi$ propagating along the long axis of the
ellipsoid.  In this model the quantization of the photon's angular momentum 
(corresponding to spin $\hbar$) arises
from an appropriately chosen of Maxwell's equations and the energy is quantized to be
$\hbar\nu$.  In a sense, not entirely clear (cf. \cite{c11}), one can think here of an
ellipsoidal soliton arising from the imposition of causality upon the solution of the
linear Maxwell equations where EM energy ${\bf E}^2+{\bf H}^2$ integrated over the
volume of the ellipsoid equals $\hbar\nu$ leading to an average intensity within the
photon-soliton  of $I_p=4\pi\hbar c^2/\gl^4$.  The word wavicle is also used in
\cite{h10}.  For a wave traveling with the speed of light parallel to the z-axis
the solution of Maxwell's equations can be any function of $z-ct$ and if
monochromatic one has a term $S(z-ct)=exp[2\pi i(z-ct)/\gl]$.  Setting
$x=rCos(\phi)$ and $y=rSin(\phi)$ in the already separated d'Alembert equation then
leads to
\bq\label{5.9}
\frac{1}{\Phi(\phi)}\frac{d^2\Phi}{d\phi^2}=m^2=-\frac{1}{R(r)}\left[\frac{d^2R}{dr^2}
+\frac{1}{r}\frac{dR}{dr}\right]
\end{equation}   
where $m^2$ is the real separation constant.  The simple plane wave solutions with
$m^2=0$ are rejected here since light is observed to travel along very narrow beams
and for $m^2=1$ one has factors of $r$ or $1/r$ with angular factors $exp(\pm i\phi)$.
This corresponds to angular momentum $L_z=(\hbar/i)\pp_{\phi}$ leading to solutions
\bq\label{5.10}
\psi(r,\phi,z-ct)=(\ga r+\gb/r)(Ae^{i\phi}+Be^{-i\phi})e^{2\pi i(z-ct)/\gl}
\end{equation}
This yields then
\bq\label{5.11}
E_z=H_z=0;\,\,E_x=(\ga r+\gb/r)\left[Ae^{i\phi}+Be^{-i\phi}\right]e^{2\pi i(z-ct)/\gl}
=\mu_0cH_y;
\end{equation}
$$E_y=i(\ga r-\gb/r)\left[Ae^{i\phi}-Be^{-i\phi}\right]e^{2\pi
i(z-ct)/\gl}=-\mu_0cH_x$$
Imposing causality leads to the result that if A or B is zero then the field must be
contained within a circular ellipsoid of length $\gl$ and cross sectional diameter
$\gl/\pi$ (cf. \cite{h10}).  The amplitude is determined by integration of the energy
${\bf E}^2+{\bf H}^2$ and the $1/r$ term is then discarded to preserve the ellipsoidal
shape; there results $A^2+B^2=1$ and $\ga^2=120h\hbar c\pi^44/\gep_0\gl^6$
(in suitable units).
In addition one expects an evanescent wave decaying like $1/r$ (with $\ga=0$)
described via
\bq\label{5.12}
E_r=\frac{\gb}{r}[A+B]=\mu_0cH_{\phi};\,\,E_{\phi}=-i\frac{\gb}{r}[A-B]=-\mu_0cH_r
\end{equation}
where $\ga r=\gb/r$ for $r=\gl/2\pi$ and $\gb^2=(\gl/2\pi)^4\times 120 n\hbar
c\pi^4/(\gep_0\gl^6)$.  The evanescent wave is believed to be responsible for 
diffraction and interference and some experimental material is sketched.

\subsection{PHOTON DYNAMICS}

We sketch here from \cite{c4} where it is shown that one can define the notions of
photon density and photon current density with certain limits.  As a trivial example
think of geometrical optics where light is treated as an ensemble of point photons
moving along definite trajectories with speed c.  In the geometric limit the photon
number density and current density are perfectly well defined as are the density and
current density for any collection of point particles.  As a second example one refers
to \cite{m7} where Mandel defines an operator $n_V$ representing the number of photons
in V as the integral over V of the photon density $D_M({\bf x})={\bf A}^{\dg}({\bf
x})\cdot{\bf A}({\bf x})$ where
\bq\label{5.13}
{\bf A}({\bf x})=L^{-3/2}\sum_{{\bf k},\gl}\vec{\gep}_{{\bf k},\gl}a_{{\bf k},\gl}
e^{i({\bf k}\cdot{\bf x}-\go t)}
\end{equation}
is the so-called detection operator (here $\vec{\gep},\,\,a,$ and $\go=c{\bf k}$ are
respectively the polarization unit vector, the annihilation operator and the frequency
of a transverse photon of wave vector {\bf k} and polarization $\gl\,\,(=1,2)$ with
$L^3$ the quantization volume).  It is shown that when the linear dimensions of V are
large compared to the photon wavelengths this definitiion of $n_V$ yields a simple for
the probability $p_V(n)$ that n photons are present in V.  It was later shown by  
Amrein \cite{a6} that $n_V$ agrees with that derived from the theory of \cite{j2}
and this all has motivated the study in \cite{c4} that a coarse grained photon density
operator can exist even though a fine grained or microscopic one may not.
\\[3mm]\indent
Thus one derives a photon density $D({\bf x})$ and a photon current density
${\bf C}({\bf x})$ to satisfy $(\bullet)\,\,\pp_tD+\na\cdot{\bf C}=0$
(conservation of photons ignoring absorption and emission).  These will be defined in
terms of vector field operators $\vec{\psi}({\bf x})$ and $\vec{\phi}({\bf x})$ which
will be referred to as the photon field (cf. here also Section 2 again).  For a volume
V large compared to the photon wavelengths $D({\bf x})$ will correctly predict
the number statistics of photons in that volume while for a time interval $[t,t+T]$
long compared to $\gl/c$ ${\bf C}({\bf x})$ correctly predicts the statistics of the
number of photons that cross the surface S in time T.  This was worked out in the
first paper of \cite{c5} for a discrete situation and is redone in the second paper
in a continuum context; we sketch this here for the free field case and, following
\cite{c5}, show that photon dynamics is a relativistically covariant theory.
Thus write
\bq\label{5.14}
\vec{\psi}({\bf x},t)=\frac{1}{\sqrt{2(2\pi)^3}}\sum_{\gl=1}^2\int
d^3k\vec{\gep}_{\gl}({\bf k})a_{\gl}({\bf k})e^{i({\bf k}\cdot{\bf x}-\go t)};
\end{equation}
$$\vec{\phi}({\bf x},t)=\frac{1}{\sqrt{2(2\pi)^3}}\sum_{\gl}\int d^3k
\left(\frac{{\bf k}}{k}\times\vec{\gep}_{\gl}({\bf k})\right)a_{\gl}({\bf k})e^{i({\bf
k}\cdot{\bf x}-\go t)}$$
where $a_{\gl}({\bf k})$ is the annihilation operator and $\vec{\gep}_{\gl}({\bf k})$
the polarization vector of a transverse photon of wave vector {\bf k} and polarization
$\gl\,\,(=1,2)$.  Evidently we have the free field equations
\bq\label{5.15}
\na\cdot\vec{\psi}=0;\,\,\na\cdot\vec{\phi}=0;\,\,\na\times\vec{\psi}+
\frac{1}{c}\pp_t\vec{\phi}=0;\,\,\na\times\vec{\phi}-\frac{1}{c}\pp_t\vec{\psi}=0
\end{equation}
Then one defines
\bq\label{5.16}
D=\vec{\psi}^{\dg}\cdot\vec{\psi}+\vec{\phi}^{\dg}\cdot\vec{\phi};\,\,
{\bf C}=c(\vec{\psi}^{\dg}\times\vec{\phi}-\vec{\phi}^{\dg}\times\vec{\psi})
\end{equation}
Evidently $(\bullet)\,\,\pp_tD+\na\cdot{\bf C}=0$ as required.  Note that D is a
positive definite operator whose integral over all space is the usual photon number 
operator
\bq\label{5.17}
\int d^3xD({\bf x}=\sum_{\gl}\int d^3ka_{\gl}^{\dg}({\bf k})a_{\gl}({\bf k})
\end{equation}
The interpretation of {\bf C} as the photon current density is justified by $(\bullet)$
and by a calculation showing that the integral of the inward normal component of {\bf
C} over the surface of an ideal photon detector equals the counting rate of the
detector (cf. \cite{c5} first paper).  The operators for the number of photons in V
and the number of photons crossing a given surface in the time interval $[t,t+T]$ are
\bq\label{5.18}
n_V=\int_Vd^3xD({\bf x});\,\,n_T=\int_t^{t+T}dt'\int_Sda{\bf n}\cdot{\bf C}({\bf x},t)
\end{equation}
({\bf n} is the unit normal to S in the direction of interest).  For a volume V large
as described the probability that V contains m photons is $(\bullet\bullet)\,\,
p_V(m)=Tr[\rho:n_V^mexp(-n_V):]/m!$ where $\rho$ is the density operator of the
radiation field and $:\,\,:$ means normal ordering.  Similarly for sufficently large T
the probability that m photons cross the surface S in time T is $(\bl\bl)\,\,
p_T(m)=Tr[\rho:n_T^mexp(-n_T):]/m!$ where S is the sensitive surface of an ideal
photon detector (with one unit quantum efficiency) and $p_T(m)$ is the photon count
distribution measured by the detector.  For these calculations see the first paper of 
\cite{c5} and \cite{m7} ($n_V$ and $n_T$ are treated as number operators).
\\[3mm]\indent
Now the transverse EM field operators ${\bf E}={\bf E}^{+}+{\bf E}^{-}$ and 
${\bf B}={\bf B}^{+}+{\bf B}^{-}$ can be expressed via
\bq\label{5.19}
{\bf E}=\frac{i}{2\pi}\sum_{\gl}\int d^3k(\hbar\go)^{1/2}\vec{\gep}({\bf
k})a_{\gl}({\bf k})e^{i({\bf k}\cdot{\bf x}-\go t)};
\end{equation}
$${\bf B}=\frac{i}{2\pi}\sum_{\gl}\int d^3k(\hbar\go)^{1/2}\left(\frac{{\bf
k}}{k}\times\vec{\gep}({\bf k})\right)a_{\gl}({\bf k})e^{i({\bf k}\cdot{\bf x}-\go
t)}$$ 
and ${\bf E}^{-}=({\bf E}^{+})^{\dg}$ with ${\bf B}^{-}=({\bf B}^{+})^{\dg}$.
One sees that the photon field vectors in \eqref{5.14} are obtained from ${\bf E}^{+}$
and ${\bf B}^{+}$ by multiplying the momentum components of ${\bf E}^{+}$ and ${\bf
B}^{+}$ by $-i[4\pi\hbar\go({\bf k})]^{-1/2}$ which corresponds to a convolution in 
position space
\bq\label{5.20}
\vec{\psi}({\bf x},t)=\int d^3yg({\bf x}-{\bf y}){\bf E}^{+}({\bf
y},t);\,\,\vec{\phi}({\bf x},t)=\int d^3yg({\bf x}-{\bf y}){\bf B}^{+}({\bf y},t)
\end{equation}
where
\bq\label{5.21}
g({\bf x})=\frac{-i}{(2\pi)^3}\int d^3k(4\pi\hbar\go)^{-1/2}e^{-i{\bf k}\cdot{\bf
x}};\,\,
\int d^3yg^{-1}({\bf x}-{\bf y})g({\bf y}-{\bf z}=\gd^3({\bf x}-{\bf z});
\end{equation}
$$g^{-1}({\bf x})=\frac{-i}{(2\pi)^3}\int d^3k(4\pi\hbar\go)^{1/2}e^{-i\bf
k}\cdot{\bf x}$$ 
One has also
\bq\label{5.22}
{\bf E}^{+}=\int d^3yg^{-1}({\bf x}-{\bf y})\vec{\psi}({\bf y},t);\,\,{\bf B}^{+}=
\int d^3yg^{-1}({\bf x}-{\bf y})\vec{\phi}({\bf y},t)
\end{equation}
It is convenient now to express the photon field as a matrix
\bq\label{5.23}
\psi_{\mu\nu}=\left(\begin{array}{cccc}
0 & -\psi_1 & -\psi_2 & -\psi_3\\
\psi_1 & 0 & \phi_3 & -\phi_2\\
\psi_2 & -\phi_3 & 0 & \phi_1\\
\psi_3 & \phi_2 & -\phi_1 & 0
\end{array}\right)
\end{equation}
Note here the similarity to \eqref{2.2} in Section 2 (apparently A. de la Torre 
was unaware of Cook's work). 
An immediate relation to the EM
field strength tensor is exhibited via
\bq\label{5.24}
F^{+}_{\mu\nu}=\left(\begin{array}{cccc}
0 & -E_1^{+} & -E_2^{+} & -E_3^{+}\\
E_1^{+} & 0 & B_3^{+} & -B_2^{+}\\
E_2^{+} & -B_3^{+} & 0 & B_1^{+}\\
E_3^{+} & B_2^{+} & -B_1^{+} & 0
\end{array}\right)
\end{equation}
One is using coordinates $x^{\mu}=(ct,x,y,z)$ and metric $g^{00}=1,\,\,g^{ii}=-1$ and
will raise and lower indices with $g^{\ga\gb}$or $g_{\ga\gb}$ as in $\psi_{\mu\nu}$
were a tensor (it turns out to transform as a tensor under displacements and spatial
rotations but not for boosts - cf. \cite{c5}).  Now define
\bq\label{5.25}
G(x)=\frac{-i}{(2\pi)^4}\int d^4k[4\pi\hbar\go(k)]^{-1/2}e^{ikx}
\end{equation}
where $x\sim x^{\mu},\,\,k\sim k^{\mu}=(k^0,{\bf k}),\,\,kx\sim k^{\mu}x_{\mu},$ and
$\go(k)=c|{\bf k}|$.  Clearly G has an inverse with $\int d^4yG^{-1}(x-y)G(y-z)=
\gd^4(x-z)$ etc.  Then one can write
\bq\label{5.26}
\psi_{\mu\nu}(x)=\int d^4y G(x-y)F^{+}_{\mu\nu}(y);\,\,F^{+}_{\mu\nu}(x)=
\int d^4yG^{-1}(x-y)\psi_{\mu\nu}(y)
\end{equation}
To see that these are equivalent to \eqref{5.20} and \eqref{5.22} note that since
$\go(k)=c|{\bf k}|$ does not depend on $k^0$ the $k^0$ integrals can be evaluated 
immediately to give $G(x)=\gd(x^0)g({\bf x})$ etc.  Although \eqref{5.20} and
\eqref{5.22} were derived from \eqref{5.14} and \eqref{5.19} for transverse photon and
EM fields one assumes that they and hence \eqref{5.26} remain valid when the EM
fields have a longitudinal component (for the free field this is of no concern).  Now
one shows that the photon field equations \eqref{5.22} are a direct consequence of the
free field Maxwell equations
\bq\label{5.27}
\pp^{\nu}F_{\mu\nu}^{+}=0;\,\,\pp_{\ga}F^{+}_{\gb\gag}+\pp_{\gb}F^{+}_{\gag\ga}+
\pp_{\gag}F^{+}_{\ga\gb}=0
\end{equation}
The second of these equations is a general operator relation following from 
$F^{+}_{\mu\nu}=\pp_{\nu}A_{\mu}^{+}-\pp_{\mu}A_{\nu}^{+}$ while the first equation
is valid in the sense that $\pp^{\nu}F^{+}_{\mu\nu}|\,>=0$ for all physically
admissable states $|\,>$ (which are defined as those satisfying the Gupta-Bleuler 
condition $(\bgs)\,\,\pp^{\mu}A_{\mu}^{+}|\,>=0$ - note the free field vector
potential satisfies the wave equation $\pp^{\nu}\pp_{\nu}A^{+}_{\mu}=0$).  Now consider
\bq\label{5.28}
\frac{\pp\psi_{\mu\nu}(x)}{\pp x^{\ga}}=\int d^4y\frac{\pp G(x-y)}{\pp
x^{\ga}}F^{+}_{\mu\nu}(y)=
\end{equation}
$$=-\int d^4y\frac{\pp G(x-y)}{\pp y^{\ga}}F^{+}_{\mu\nu}(y)=\int d^4y G(x-y)\frac
{\pp F^{+}_{\mu\nu}(y)}{\pp y^{\ga}}$$
Neglect of the integrated part is justified via $G(x)\to 0$ for $x\to\infty$ and 
$F^{+}_{\mu\nu}(x)\to 0$ at spatial infinity.  From \eqref{5.27} one has then
\bq\label{5.29}
\pp^{\nu}\psi_{\mu\nu}=0;\,\,\pp_{\ga}\psi_{\gb\gag}+\pp_{\gb}\psi_{\gag\ga}+
\pp_{\gag}\psi_{\ga\gb}=0
\end{equation}
and these are equivalent to the original photon field equations\eqref{5.15}; again
one restricts to the subspace of physical states.  Although these have the appearance
of tensor equations they are not manifestly covariant since $\psi_{\mu\nu}$ is not a
tensor (cf. Section 2 for comments in this direction).  Nevertheless the equations are
shown to be invariant under Lorentz transformations because the photon field
$\psi_{\mu\nu}$ is defined in terms of the tensor $F^{+}_{\mu\nu}$ in the same way in
each Lorentz frame (see here \cite{c5} for details).
\\[3mm]\indent
Finally one considers the matrix of Hermitian operators
\bq\label{5.30}
N^{\ga\gb}=\psi_{\gl}^{\dg\ga}\psi^{\gl\gb}+\psi^{\dg\gl\gb}\psi_{\gl}^{\ga}+
\frac{1}{2}g^{\ga\gb}\psi^{\dg\mu\nu}\psi_{\mu\nu}
\end{equation}
with is analogous to the EM energy momentum tensor.  One checks easily that
$(\clubsuit\clubsuit)\,\,\pp_{\gb}N^{\ga\gb}=0\Rightarrow M^{\ga}=\int d^3xN^{\ga 0}$
is conserved as the photon field develops in time.  In fact one can write
\bq\label{5.31}
N^{\ga\gb}=\left(\begin{array}{cccc}
D & C_1/c & C_2/c & C_3/c\\
C_1/c & S_{11}/c^2 & S_{12}/c^2 & C_{13}/c^2\\
C_2/c & S_{21}/c^2 & S_{22}/c^2 & S_{23}/c^2\\
C_3/c & S_{31}/c^2 & S_{32}/c^2 & S_{33}/c^2
\end{array}\right)
\end{equation}
Here
$(\spadesuit\spadesuit)\,\,S_{ij}=c^2[D\gd_{ij}-(\psi_i^{\dg}\psi_j+\psi^{\dg}_j\psi_i)
-(\phi^{\dg}_i\phi_j+\phi^{\dg}_j\phi_i)]$ is a $3\times 3$ matrix analogous to the
Maxwell stress tensor and $(\clubsuit\clubsuit)$ now takes the form
\bq\label{5.32}
\pp_tD+\na\cdot{\bf C}=0;\,\,\pp_tC_i+\frac{\pp S_{ij}}{\pp x^j}=0;
\end{equation}
$$M^0=\int d^3x
D({\bf x})=const.;\,\,cM^i=\int d^3x C_i({\bf x}=const.$$
These equations express the local and global conservation of photons.  One shows
that $N^{\ga\gb}$ does not transform as a tensor (except for coordinate displacements
and spatial rotations) and in fact there is no general transformation law relating the
components of $N^{\ga\gb}$ in different Lorentz frames.  Nevertheless \eqref{5.32}
are covariant since the photon field equations are covariant and $N^{\ga\gb}$ is
constructed from the photon field in the same way in each frame.  In particular the
number of photons $M^0$ is independent of time in each Lorentz frame and considerable
calculation also shows that $M^0$ is a scalar (using the condition $(\bgs)$).

\section{SOME SPECULATIONS ON THE AETHER}
\renewcommand{\theequation}{6.\arabic{equation}}
\setcounter{equation}{0}

The aether has been reviewed in \cite{c1,c6} to a certain extent and in \cite{c1}
some speculations were advanced concerning a possible geometry for the aether.  These
were based on work of \cite{a1,a2,b1,c1,c8,c10,c7,c9,c6,f1,f2,i2,v1} and we sketch
here some variations and embellishments.  First we note from \eqref{5.15} that the
components $\psi_i$ and $\phi_i$ satisfy the massless KG equation so for analysis of
photons one needs 6 components $(\psi_i,\phi_i)$ each satisfying a massless KG
equation.  However the equations \eqref{5.15} are exactly the same as the Maxwell equations
\eqref{3.7} so one could also imagine introducing a vector $\Psi=(-{\bf A},\phi)$ with
$A_{\mu\nu}=\Psi_{\mu,\nu}-\Psi_{\nu,\mu}$ to generate the photon equations for a free
field with $\bx\Psi=0$ (see e.g. \cite{r6}).  In this spirit then one would have
a 4-vector $\Psi$ satisfying the massless KG equation to serve as a generator of
photon activity.  In any event we will think of fields labeled $\psi_i$ for $i=0,1,2,3$
as characterizing photon dynamics with each component satisfying the massless KG equation.
Then we will apply the machinery of $(x,\psi)$ duality of Faraggi-Matone and Vancea
(see especially \cite{f1,v1}) to express the coordinates $x^{\mu}$ in terms of the
fields $\psi_i$ arising from $\Psi$ (which will be called aether fields); they are seen to be
``potential" fields for the photon fields $\psi_i,\,\,\phi_i$ of Section 5.1.
\\[3mm]\indent
As background here we refer to a lovely paper 
\cite{i2} of P. Isaev where he makes
conjectures, with supporting arguments, which arrive at a definition of the aether as a
Bose-Einstein condensate of neutrino-antineutrino pairs of Cooper type
(Bose-Einstein condensates of various types have been considered by others in this
context - cf. \cite{c15,e1,j3}).  The equation for
the
$\psi$-aether is then a  solution of the massless Klein-Gordon (KG) equation (photon
equation)
$(\hbar^2\gD-(\hbar^2/c^2)\pp_t^2)\psi=0$.
This $\psi$ field heuristically acts as a carrier of
waves (playground for waves) and one might say that special relativity (SR) 
is a way of including the influence of the aether on physical processes and consequently
SR does not see the aether
(cf. here also the idea of a Dirac
aether in
\cite{c16,c17,d3,o1} and Einstein-aether theories as in \cite{e1,j3} - this is
discussed further in \cite{c1}). 
In the electromagnetic (EM) theory in \cite{i2} one looks at
$\vec{\psi}=(\phi,\vec{A})$ with
$\bx\psi_i=0$ as the defining equation for a real $\psi$-aether, in terms of the
potentials $\phi$ and $\vec{A}$ which therefore define the $\psi$-aether.  EM waves are
then considered as oscillations of the
$\psi$ aether and wave processes in the aether accompanying a moving
particle determine wave properties of the particle.  Interesting examples involving standing EM
waves in a spherical resonator are attributed to oscillations of the $\psi$ aether  and 
references to superconductivity \`a la Volovik \cite{v2,v3} are indicated.
\\[3mm]\indent
In \cite{f1} Faraggi and Matone develop a theory of $x-\psi$
duality, related to Seiberg-Witten theory in the string arena, which
was expanded in various ways in
\cite{a1,a2,b1,c8,c10,c7,m8,v1}. Here one works from a
stationary SE $[-(\hbar^2/2m)\gD +V(x)]\psi=E\psi$, and, assuming for
convenience one space dimension, the space variable $x$ is determined
by the wave function $\psi$ from a prepotential ${\mf F}$ via
Legendre transformations.  The theory suggests that
$x$ plays the role of a macroscopic variable for a statistical system
with a scaling term involving $\hbar$.  Thus define a prepotential ${\mf
F}_E(\psi)={\mf F}(\psi)$ such that the dual variable
$\psi^D=\pp{\mf F}/\pp\psi$ is a (linearly independent) solution of
the same SE.  Take V and E real so that
$\bar{\psi}=\psi^D$ qualifies and write $\pp_x{\mf
F}=\psi^D\pp_x\psi=(1/2)[\pp_x(\psi\psi^D)+W)]$ where W is the
Wronskian.  This leads to ($\psi^D=\bar{\psi}$) the relation ${\mf
F}=(1/2)\psi\bar{\psi}+(W/2)x$ (setting the integration constant to
zero). Consequently, scaling W to $-2i\sqrt{2m}/\hbar$ one obtains
\bq\label{6.1}
\frac{i\sqrt{2m}}{\hbar}x=\frac{1}{2}\psi\frac{\pp{\mf
F}}{\pp\psi}-{\mf F}\equiv
\frac{i\sqrt{2m}}{\hbar}x=\psi^2\frac{\pp{\mf F}}{\pp\psi^2}-{\mf F}
\end{equation} 
which exhibits $x$ as a Legendre transform of ${\mf
F}$ with respect to
$\psi^2$.  Duality of the Legendre transform then gives also 
\bq\label{6.2} 
{\mf
F}=\phi\pp_{\phi}\left(\frac{i\sqrt{2m}\,x}{\hbar}\right)-
\left(\frac{i\sqrt{2m}\,x}{\hbar}\right);\,\,
\phi=\pp_{\psi^2}{\mf F}=\frac{\bar{\psi}}{2\psi}
\end{equation} 
so that ${\mf F}$ and $(i\sqrt{2m}\,x/\hbar)$ form a
Legendre pair.  In particular one has 
$\rho=|\psi|^2=\frac{2i\sqrt{2m}}{\hbar}x+2{\mf F}$ which also
relates ${\mf F}$ and the probability density.
In any event one sees that the wave function $\psi$ specifically
determines the location of the ``particle" whose quantum
evolution is described by 
$\psi$.  We mention here also that the (stationary) SE can be
replaced by a third order equation
\bq\label{6.3} 
4{\mf F}'''+(V(x)-E)({\mf F}'-\psi{\mf F}'')^3=0;\,\,{\mf
F}'\sim\frac{\pp{\mf F}}{\pp
\psi}
\end{equation} 
and a dual stationary SE has the form
\bq\label{6.4}
\frac{\hbar^2}{2m}\frac{\pp^2x}{\pp\psi^2}=\psi[E-V]\left(\frac{\pp
x}{\pp\psi}\right)^3
\end{equation}
A noncommutative version of this is developed in the second paper of \cite{v1}.
\\[3mm]\indent
We mention
\cite{m1,p1,p2} for some material on the aether and the vacuum and refer to  
the bibliography for other references.  We
sketch first some material from \cite{a1,a2,b1,v1} which
extends the SE theory to the Klein-Gordon (KG) equation.  Following \cite{v1} take
a spacetime manifold M with a metric field $g$ and a scalar field $\psi$
satisfying the KG equation.  Locally one has cartesian coordinates
$x^{\ga}\,\,(\ga=0,1.\cdots,n-1)$ in which the metric is diagonal
with $g_{\ga\gb}(x)=\eta_{\ga\gb}(x)$ and the KG equation has the form
$(\bx_x+m^2)\psi(x)=0$ ($\bx_x\sim (\hbar^2/c^2)[(\pp_t^2/c^2)-\na^2]$). 
Defining prepotentials such that
$\tl{\psi}^{(\ga)}=\pp{\mf F}^{(\ga)}[\psi^{(\ga)}]/\pp\psi^{(\ga)}$
where $\psi^{(\ga)}$ and $\tl{\psi}^{(\ga)}$ are two linearly
independent solutions of the KG equation depending on parameters
$x^{\ga}$ one has as above (with a different scaling factor)
\bq\label{6.5}
\frac{\sqrt{2m}}{\hbar}x^{\ga}=\frac{1}{2}\psi^{(\ga)}\frac{\pp{\mf
F}^{(\ga)}[\psi^{\ga)}]}{\pp\psi^{(\ga)}}-{\mf
F}^{(\ga)};\,\,[\pp^{\ga}\pp_{\ga}-V^{\ga}]\psi^{\ga}=0
\end{equation}
This is suggested in \cite{f1} and used in \cite{v1}; the factor $\sqrt{2m}/\hbar$ is simply
a scaling factor (possibly too stringent here) and it would be more productive to scale
$x^0\sim ct$ differently or in fact to scale all variables as indicated in \cite{c1}
(cf. below for a general scaling).
Locally ${\mf F}^{(\ga)}$ satisfies the third order equation
\bq\label{6.6}
4{\mf F}^{(\ga)^{'''}} +[V^{(\ga)}(x^{\ga})+m^2](\psi^{(\ga)}{\mf F}^{(\ga)^{''}}-{\mf
F}^{(\ga)')})^3=0
\end{equation}
where $'\sim\pp/\pp\phi^{(\ga)}$ and a (quantum) potential $V^{\ga}$ has the form
\bq\label{6.7}
\left.V^{(\ga)}(x^{\ga})=\left[\frac{1}{\psi(x)}
\sum_{\gb=0,\,\gb\ne\ga}^{n-1}\pp^{\gb}\pp_{\gb}
\psi(x)\right]\right|_{x^{\gb\ne\ga}\,fixed}
\end{equation}
\indent
We go back to \cite{f1} now and derive equations for the KG equation with $m=0$ from the
beginning (rather than rescaling and then taking $m\to 0$).  Further we proceed with more
detail and show how a general scaling will involve insertion of some variable factors
(cf. also \cite{c8,c9} for various scaling factors).  Thus consider $(1/c^2)\psi_{tt}-
\gD\psi=0$ with $x^0=ct$ and write out explicitly ($i=1,2,3$)
\bq\label{6.8} 
\frac{1}{c^2}\pp_t^2\psi^0-V^0\psi^0=0;\,\,V^0=\frac{\gD\psi}{\psi}=
\frac{(1/c^2)\psi_{tt}}{\psi};
\end{equation}
$$\pp_i^2\psi^i-V^i\psi^i=0;\,\,V^i=\frac{\left(\frac{1}{c^2}\pp_t^2\psi-\sum_{j\ne i}
\pp_j^2\psi\right)}{\psi}$$
Here $V^i$ is thought of as $V^i(x^i)$ (where in fact $V^i=V^i(x^i,x^j,x^0)$ with $j\ne i$
and
$x^0,\,x^j$ are considered as parameters).  Similarly $V^0=V^0(x^0)$ ($\equiv V^0(x^0,x^i)$).
Now e.g. for $\psi^0$ and $\tl{\psi}^0$ linearly independent solutions of the first equation
in \eqref{6.8} one has $\psi_{tt}^0\tl{\psi}^0=\psi^0\tl{\psi}^0_{tt}$ which implies
\bq\label{6.9}
W^0(t)=(\psi^0\tl{\psi}^0_t(t)-\tl{\psi}^0\psi_t^0)(t)=2c\gag(x^i)
\end{equation}
Here, as specified above, $\tl{\psi}^0=\pp{\mf F}^0/\pp\psi^0$, and
\bq\label{6.10} 
\pp_t{\mf F}^0={\mf F}^0_{\psi}\psi_t=\tl{\psi}^0\psi_t\Rightarrow
\end{equation}
$$\Rightarrow \frac{1}{2}\pp_t(\psi^0\tl{\psi}^0)-\tl{\psi}^0\psi^0_t=\frac{1}{2}
(\psi^0\tl{\psi}^0_t-\psi_t^0\tl{\psi}^0)=\frac{1}{2}W^0=c\gag(x^i)$$
and consequently one can write
\bq\label{6.11}
c\gag(x^i)t=\frac{1}{2}\psi^0\frac{\pp{\mf F}^0}{\pp\psi^0}-{\mf F}^0={\mf E}^0
\end{equation}
This leads to (for $\psi^0\sim\phi$)
\bq\label{6.12}
c\gag(x^i)=\frac{\pp{\mf E}^0}{\pp\phi}\frac{d\phi}{dt}=\left[\frac{1}{2}\left({\mf
F}^0_{\phi} +\phi\frac{\pp^2{\mf F}^0}{\pp\phi^2}\right)-{\mf
F}^0_{\phi}\right]\frac{d\phi}{dt}=
\end{equation}
$$=\frac{1}{2}\left(\phi\frac{\pp^2{\mf F}^0}{\pp\phi^2}-{\mf
F}^0_{\phi}\right)\frac{d\phi}{dt}=\frac{1}{2}E^0\frac{d\phi}{dt}$$
Similarly we write, using \eqref{6.8},
\bq\label{6.13}
\tl{\psi}^i=\frac{\pp{\mf
F}^i}{\pp\psi^i};\,\,W^i=\psi^i\pp_t\tl{\psi}^i-\tl{\psi}^i\pp_t\psi^i;\,\,\gb^i(x^0,x^j)x^i
=\frac{1}{2}\psi^i\frac{\pp{\mf F}^i}{\pp\psi^i}-{\mf F}^i={\mf E}^i
\end{equation}
Consequently ($\psi_i\equiv\psi^i$)
\bq\label{6.14}
\gag dx^0=c\gag dt=\frac{1}{2}E^0d\psi^0;\,\,\gb^i
dx^i=\frac{1}{2}E^id\psi^i=\frac{1}{2}\left(\psi^i\frac{\pp^2{\mf F}^i}{\pp\psi_i^2}-
\frac{\pp{\mf F}^i}{\pp\psi^i}\right)d\psi^i
\end{equation}
Since $\pp_t=\pp_{\phi}(d\phi/dt)$, etc. one can write then
\bq\label{6.15}
\pp_t=\left(\frac{2c\gag}{E^0}\right)\pp_{\phi};\,\,\pp_i=\left(\frac
{2\gb}{E^i}\right)\frac{\pp}{\pp\psi^i}
\end{equation}
The extraneous variables are considered as parameters when concentrating on one $x^i$ or
$x^0$ and we note from \eqref{6.11} or \eqref{6.13} that $x^0$ or $x^i$ can be considered as a
function of $\phi=\psi^0$ or $\psi^i$ and ${\mf F}^i$ is a function of $\psi^i$ (satisfying
ordinary differential equations as in \eqref{6.6} - with $m=0$).
Here \eqref{6.14}-\eqref{6.15} 
represents an induced parametrization on the spaces $T_P(U)$ and 
$T^*_P(U)$ ($P\in U$ - local tangent and cotangent spaces).  Now using the linearity of the
metric tensor field one sees that the components of the metric in the $\{(\psi^{\ga},{\mf
F}^{\ga})\}$ parametrization are ($\gb^0=c\gag$)
\bq\label{6.16}
G_{\ga\gs}(\psi)=\frac{E^{\ga}E^{\gs}}{4\gb^{\ga}\gb^{\gs}}\eta_{\ga\gs}(x)
\end{equation}
(cf. \cite{v1}).
Now following \cite{v1} let $z^{\mu}\,\,(\mu=0,1,\cdots,n-1)$ be a general coordinate system
in U and write the coordinate transformation matrices via
\bq\label{6.17}
A_{\mu}^{\ga}=\frac{\pp x^{\ga}}{\pp z^{\mu}};\,\,(A^{-1})^{\mu}_{\ga}=\frac{\pp
z^{\mu}}{\pp x^{\ga}}
\end{equation}
The metric then takes the form
\bq\label{6.18}
g_{\mu\nu}(z)=\frac{4\gb^{\ga}\gb^{\gs}}{E^{\ga}E^{\gs}}A_{\mu}^{\ga}A_{\nu}^{\gs}
G_{\ga\gs}(\psi)
\end{equation} 
The components of the metric connection can be computed via
\bq\label{6.19}
\gG^{\rho}_{\mu\nu}=\frac{1}{2}g^{\rho\gs}(z)\sum_{{\mc P}}\gep_{{\mc
P}}{\mc P}\left[\frac{\pp g_{\gs\nu}(z)}{\pp z^{\mu}}\right]
\end{equation}
where ${\mc P}$ is a cyclic permutation of the ordered set of indices $\{\gs\nu\mu\}$
and $\gep_{{\mc P}}$ is the signature of ${\mc P}$.  Via the coordinate transformation
\eqref{6.17} the function $\psi^{\ga}$ depends on all the $z^{\mu}$.  The metric
connection \eqref{6.19} can be expressed in the $\{\psi^{\ga},{\mf F}^{\ga}\}$
parametrization and in
\cite{v1} one computes also the components of the curvature tensor, the Ricci tensor, and
the scalar curvature and gives an expression for the Einstein equations (we omit the details
here).  The same procedure apply to our formulas above which leads us to state heuristically
\begin{theorem}
The formulas \eqref{6.14}, \eqref{6.15}, \eqref{6.16}, \eqref{6.17}, \eqref{6.18}, and
\eqref{6.19}, and their continuations determine a geometry for a putative aether,
expressed in terms of our so-called aether fields $\psi_i$.
\end{theorem}
\indent
{\bf REMARK 1.1.}
These matters are taken up again in \cite{a2} for a general curved spacetime and some
sufficient constraints are isolated which make the theory work.  Also in both papers
a quantized version of the KG equation is also treated and the relevant $x-\psi$
duality is spelled out in operator form.  We omit this also in remarking that the
main feature here for our purposes is the fact that one can describe spacetime geometry
(at least locally) in terms of (field) solutions of a KG equation and prepotentials
(which are themselves functions of the fields).  In other words the coordinates are
programmed by fields and if the motion of some particle of mass m is involved then
its coordinates are choreographed by the fields with a quantum potential eventually entering
the picture via \eqref{6.7}.  In \cite{a1} a similar duality is worked out for the Dirac
field and cartesian coordinates and to connect this with the aether idea one should examine
such formulas for $m\to 0$.$\hfill\bs$
\begin{example}
One knows that general solutions of the massless KG equation will have the form $\psi=
\psi({\bf a}\cdot{\bf x}-ct)$ with $|{\bf a}|=1$.  For example take $\psi=exp(\sum
a_ix_i-ct)$
with $(1/c^2)\psi_{tt}=\psi$ and $\psi_{ii}=a_i^2\psi$.  This leads to
\bq\label{6.20}
V^0=1;\,\,V^i=1-\sum_{j\ne i}a_j^2
\end{equation}
Hence
\bq\label{6.21}
\frac{1}{c^2}\pp_t^2\psi^0-\psi^0=0;\,\,\pp_i^2\psi^i-(1-\sum_{j\ne i}a_j^2)\psi^i=0
\end{equation}
On the other hand if $\psi=f({\bf a}\cdot{\bf x}-ct)$ one gets
\bq\label{6.22}
V^0=\left(\frac{f''}{f}\right)({\bf a}\cdot{\bf x}-ct);\,\,
\,\,V^i=\left(1-\sum
_{j\ne i}a_j^2\right)\left(\frac{f''}{f}\right)({\bf a}\cdot{\bf x}-ct)
\end{equation}
Setting $f''/f=g(x^i,x^0)$ one has
\bq\label{6.23}
\pp_0^2\psi^0-g(x^i,x^0)\psi^0=0;\,\,\pp_i^2\psi^i-\left(1-\sum_{j\ne i}
a_j^2\right)g(x^i,x^j,x^0)\psi^i
\end{equation}
Here the $x^i$ or $(x^j,\,x^0)$ are considered as parameters.$\hfill\bs$
\end{example}
\begin{example}
Consider a simple situation with two $x^i$ variables and $x^0=ct$ and take
$a_1=a_2=1/\sqrt{2}$.  Then $V^0=1$ and $V^i=1-(1/2)=1/2$ with
\bq\label{6.24}
\frac{1}{c^2}\pp_t^2\psi^0=\psi^0;\,\,\frac{\pp^2\psi^i}{\pp(x^i)^2}\psi^i=\frac{1}{2}\psi^i
\end{equation}
Hence we can take
\bq\label{6.25}
\psi^0=A_0e^{ct};\,\,\psi^i=A_ie^{(1/\sqrt{2})x^i};\,\,\tl{\psi}^i=
\tl{A}_ie^{-(1/\sqrt{2})x^i};\,\,\tl{\psi}^0=\tl{A}_0e^{-ct}
\end{equation}
Now $\psi^i\tl{\psi}^i=\gk_i$ for $i=0,1,2$ so (recall $\gb^0=\gag$ and $x^0=ct$)
\bq\label{6.26}
\gb^i x^i=\frac{1}{2}\gk_i-{\mf F}^i={\mf E}^i\,\,\,(i=0,1,2)
\end{equation}
and 
\bq\label{6.27}
\frac{1}{2}E^i=\frac{\pp}{\pp\psi^i}\left(\frac{1}{2}\gk_i-{\mf F}^i\right)=-\frac{\pp{\mf
F}^i} {\pp\psi^i}=-\tl{\psi}^i\,\,\,(i=0,1,2)
\end{equation}
(the $\gb^i$ here need not depend on other variables).
Consequently one has
\bq\label{6.28}
G_{\ga\gs}(\psi)=\frac{E^{\ga}E^{\gs}}{4\gb^{\ga}\gb^{\gs}}\eta_{\ga\gs}(x)=
\frac{\tl{\psi}^{\ga}\tl{\psi}^{\gs}}{\gb^{\ga}\gb^{\gs}}\eta_{\ga\gs}(x)
\end{equation}
and this exhibits in a simple example the manner in which the metric can depend on the
fields.$\hfill\bs$
\end{example}
\begin{example}
We look now at the more complicated situation for $\psi=f({\bf a}\cdot{\bf x}-ct)$ as in
(6.22)-(6.23).  Here $f''/f=g$ could be a fairly general function with argument ${\bf
a}\cdot{\bf x}-ct$ and in the equations $\pp_i^2\psi^i=\ga_ig\psi^i$ the function $g_i$ is
considered as a function of $x^i$ with the other $x^j$ as parameters.  Let $\psi^i$ and 
$\tl{\psi}^i$ be two solutions ($i=0,1,2,3$ say) and look at ($\psi_i\equiv \psi^i$)
\bq\label{6.29}
{\mf E}^i=\gb^ix^i=\frac{1}{2}\psi^i\frac{\pp{\mf F}}{\pp\psi^i}-{\mf F}^i;\,\,
E^i=\psi^i\frac{\pp^2{\mf F}}{\pp\psi^2_i}-\frac{\pp{\mf F}}{\pp\psi^i}
\end{equation}
Recall $\tl{\psi}^i=\pp{\mf F}/\pp\psi^i$ and we can write, from Item 3 in Section 2,
$\phi^i=\pp{\mf F}/\pp(\psi^i)^2=\tl{\psi}^i/2\psi^i$ (although this will not be used here).
In terms of the two fields $\psi$ and $\tl{\psi}^i$ one has
\bq\label{6.30}
{\mf E}^i=\frac{1}{2}\psi^i\tl{\psi}^i-{\mf F}^i=\gb^ix^i;\,\,{\mf F}^i={\mf
F}^i(\psi^i,\tl{\psi}^i,x^i,\gb^i);
\end{equation}
$$E^i=\psi^i\frac{\pp\tl{\psi}^i}{\pp\psi^i}-\tl{\psi}^i;\,\,\gb^idx^i=\frac{1}{2}E^id\psi^i$$
In particular $E^i$ is expresed directly in terms of the fields $\psi^i$ and $\tl{\psi}^i$;
no extraneous variables are explicit.  Now $\psi^i$ and $\tl{\psi}^i$ are linearly
independent solutions of $\pp^2_i\psi^i=\ga_ig\psi$ but they are linked by a Wronskian
$W_i=(\pp_x\psi^i)\tl{\psi}^i-\psi^i(\pp_x\tl{\psi}^i)=-2\gb^i$ where $\gb^i$ does not
depend on $x^i$ (only perhaps on the other $x^j$).  One can write now
\bq\label{6.31}
\pp_x\left(\frac{\tl{\psi^i}}{\psi^i}\right)=\frac{W_i}{\psi_i^2}\Rightarrow
\tl{\psi}^i=\psi^i\int^x\frac{W_idx}{\psi_i^2}+c\psi^i
\end{equation}
Formally this suggests
\bq\label{6.32}
\frac{\pp\tl{\psi}^i}{\pp\psi^i}=-2\gb^i\int^x\frac{dx}{\psi_i^2}+
4\gb^i\psi^i\int^x\frac{dx}{\psi_i^3}+c
\end{equation}
from which follows 
\bq\label{6.33}
E^i=\psi^i\left[-2\gb^i\int^x\frac{dx^i}{\psi_i^2}+c+4\gb^i\psi^i\int^x\frac{dx^i}{\psi_i^3}-
\frac{2\gb^i}{\psi^i}\frac{dx^i}{d\psi^i}\right]+
\end{equation}
$$+2\gb^i\psi^i\int^x\frac{dx^i}{d\psi^i}-c\psi^i=4\gb^i\psi_i^2\int^x\frac{dx^i}{\psi_i^3}-
2\gb^i\frac{E^i}{2\gb^i}\Rightarrow E^i=2\gb^i\psi_i^2\int^x\frac{dx^i}{\psi_i^3}$$
Thus $E^i$ can be expressed entirely in terms of the field $\psi^i$. 
$\hfill\bs$
\end{example}
\indent
One notes here that these arguments and results hold for any $\psi^{\ga},\,V^{\ga}$ as in
(6.5)-(6.7) so we state heuristically
\begin{theorem}
The objects $E^{\ga}$ used in constructing the geometry can be expressed in terms of fields
$\psi^{\ga}$ as in (6.34).
\end{theorem}

\end{document}